\documentclass[10pt,twoside,preprint]{elsarticle}

\usepackage{amssymb}
\usepackage{amsmath}
\usepackage{subfigure}
\usepackage{graphicx}
\usepackage{caption2}

\usepackage{algorithm} 
\usepackage{algorithmic} 
\usepackage{multirow} 
\usepackage{amsmath}
\usepackage{xcolor}
\usepackage{float}

\usepackage{amssymb}
\usepackage{amsthm}

\theoremstyle{definition}

\theoremstyle{remark}

\usepackage{geometry}   
\usepackage[]{caption2}

\begin{document}
\begin{frontmatter}

\title{Predicting the evolution of complex networks via local information}


\author[rvt,focal,oth]{Tao Wu\corref{cor1}}
\ead{wutaoadeny@gmail.com}
\author[rvt,focal,oth]{Leiting Chen}
\cortext[cor1]{Corresponding author}


\address[rvt]{Department of Computer Science and Engineering, University of Electronic Science and Technology of China}
\address[focal]{Institute of Electronic and Information Engineering in Dongguan, University of Electronic Science and Technology of China}
\address[oth]{Digital Media Technology Key Laboratory of Sichuan Province}

\begin{abstract}
Almost all real-world networks are subject to constant evolution,
and plenty of evolving networks have been investigated to uncover
the underlying mechanisms for a deeper understanding of the
organization and development of them. Compared with the rapid
expansion of the empirical studies about evolution mechanisms
exploration, the future links prediction methods corresponding to
the evolution mechanisms are deficient. Real-world information
always contain hints of what would happen next, which is also the
case in the observed evolving networks. In this paper, we firstly
propose a structured-dependent index to strengthen the robustness of
link prediction methods. Then we treat the observed links and their
timestamps in evolving networks as known information. We envision
evolving networks as dynamic systems and model the evolutionary
dynamics of nodes similarity. Based on the iterative updating of
nodes' network position, the potential trend of evolving networks is
uncovered, which improves the accuracy of future links prediction.
Experiments on various real-world networks show that the proposed
index performs better than baseline methods and the spatial-temporal
position drift model performs well in real-world evolving networks.
\end{abstract}


\end{frontmatter}

\section{Introduction}\indent

Complex networks involve the characterization of complex systems of
interacting entities, where nodes denote the entities and links
represent the relationships or interactions between them. Complex
networks provide an ideal tool for understanding the structure and
function of many complex systems in nature, technology and society
[1]. With great brilliance, various kinds of methods have been
proposed to address the network analysis problem, including
community detection [2, 3], cascades prediction [4, 5], network
controllability [6], etc. In particular, the evolution trend, which
describe the grow and change of network topology through the
addition or deletion of nodes and links, is a critical aspect of
network analysis. Up to now, the evolution mechanisms that have been
preliminarily verified include preference attachment [7], small
world [8], homophily [9], clustering [10], etc. Since uncovering how
real-world networks are evolved can considerably deepen the
understanding about many complex systems and guide the behavior of
individuals, the question which we will address here is how to
predict the evolution trend of complex networks without losing
generality. According to Ref. [11], an effective link prediction
algorithm provides strong evidence of the corresponding mechanism(s)
of network organization. Here, we will in fact answer the question
by developing a structure-dependent link prediction index for
ranking the candidates of future links and an explicit
spatial-temporal network position drift model for uncovering the
potential trend of network evolution.

In recent years, link prediction problem has attracted a
considerable amount of attention in complex network studies [12-15].
Especially, D. Liben-Nowell et al. [15] argue that the link
prediction problem asks to what extent can the evolution of a
complex network be modeled using features intrinsic to the network
itself ? They summarize many similarity indices based on network
structure and find that there is indeed useful information contained
in the network topology alone by comparing them with random
predictors. Commonly, the topology-based methods can be divided into
three classes.  The first class is similarity-based methods, which
assume that links between more similar nodes are of higher existing
likelihood, and the similarity of the two endpoints can be
transferred through the links. The similarity-based methods can be
subdivided into neighbor-based methods and distance-based methods.
Neighbor-based methods are based on the idea that two nodes are more
likely to generate a link in the near future if they have more
common neighbors, such as Sorensen [16], LHN [17], CN[18], etc.
Distance-based methods suppose that link probability is determined
by distance or number of the shortest path between nodes, such as LP
[19], Katz [20], LHN-II [17], etc.  The second class of link
prediction methods is maximum likelihood estimation methods. Two
popular methods of this type are the hierarchical structure model
(HSM) [21] and the stochastic block model (SBM) [22]. The third
class of link prediction methods is machine learning based methods.
The main methods of this type are the supervised learning method
[23] and negative matrix factorization (NMF) mehtod [24]. Owning to
its simplicity, the study on similarity-based algorithms is the
mainstream issue.

However, in most of the existing works, a typical method is to
consider an algorithm's accuracy in reproducing known network links
that have been removed from a test dataset. An accurate prediction,
however, is not necessarily a useful one. Specially speaking, the
distinction between future links prediction and missing links
prediction is actually never that clear, while Ref. [25] offers
evidence that missing links are more likely to be the links
connecting low-degree nodes. As almost all real-world networks are
evolving all the time [26], one has to investigate the evolution
trend of networks, i.e., what are the future links of evolving
networks. This is not a trivial problem. It has already been pointed
out that many new members are the neighbors or second level
neighbors of the current group members and they are more likely to
join the group with the clustering of the group increases from a
group-level perspective in evolving real-world networks [27, 28].
That is to say, node groups with higher clustering level are more
likely attract nodes than the uncompacted ones. Furthermore, the
result of Ref. [29] shows that the auto-correlation function of the
successive states of communities is continuous, which indicates that
the current states of networks are associated with the states at the
last time point. Next the Ref. [30] recognizes a pattern that links
are highly biased towards young nodes. Thus the timestamps of
network interactions signified by network links has the potential to
hint the trend of network evolution. However, how all these factors
influence the future links prediction still remain unclear.
Therefore, it is necessary to explore the problem of evolving
networks prediction in combination with these factors.

In this paper, in order to adapt to networks with different
structure properties, we propose a structure-dependent index
combining the virtues of neighbor-based methods and distance-based
methods. Moreover, unlike the existing link prediction methods which
assume the similarities of nodes are constant, we envision evolving
networks as dynamic systems and investigate the evolutionary
dynamics of similarities. The structure features of nodes'
neighborhood and the timestamps of network interactions are called
spatial factor and temporal factor respectively. Combining the
spatial and temporal factors, we propose a dynamic network position
drift model, in which node's network position undergoes migration
and variation under the influence of its neighbors' attractiveness.
The variation of node's network position reflects the potential
trend underlying the current network structure, and the iterative
updating of node's network position would lead to an evolved network
representing the similarity relationships of network nodes in the
near future. Finally, according to the empirical results on five
real-world networks, we find that the structure-dependent index is
effective and robust for link prediction and the spatial-temporal
position drift model has satisfactory performance.

The rest of the paper is organized as follows. Section 2 introduces
some indices as baselines. Section 3 presents the
structure-dependent index and the spatial-temporal position drift
model. Section 4 gives the empirical analysis. Discussion and
conclusion are drawn in Section 5.

\section{Preliminaries}\indent

\subsection{Problem and evaluation}\indent

Consider an undirected simple network $ G = (V, E) $, where $ V =
\{v_i\} $ is the set of nodes and $ E $ is the set of links.
Multiple links and self-connections are not allowed. Each link $e =
(u,v,t) \in E$ represents an interaction between $u$ and $v$ that
took place at time $t$. Let $ G[t,t']$ denote the subgraph of $ G $
consisting of all links with timestamps between $t$ and $t'$. We
choose three timestamps $ {\rm{t}}_0 {\rm{ < t}}_{\rm{1}} {\rm{ <
t}}_{\rm{2}}$ and get subgraph ${\rm{G[t}}_{\rm{0}} {\rm{,
t}}_{\rm{1}} {\rm{]}}$ and ${\rm{G[t}}_1 {\rm{, t}}_2 {\rm{]}}$. We
refer to ${\rm{[t}}_0 {\rm{, t}}_1 {\rm{]}}$ as the training
interval and ${\rm{[t}}_1 {\rm{, t}}_2 {\rm{]}}$ as the probe
interval. Commonly, networks grow through the addition of nodes and
links, and it is not sensible to predict the links of $G[t_1, t_2]$
whose endpoints are not present in $G[t_0, t_1]$. Thus we eliminate
the new added nodes within $[t_1, t_2]$. We define $ET$ and $EP$ to
denote the set of links in $G[t_0, t_1]$ and $G[t_1, t_2]$
respectively. Clearly, we have $ET \cap EP = \emptyset$. We use $U$
to denote the universal set containing all $|V'|(|V'|-1)/2$ possible
links, where $V'$ denotes the set of nodes in $G[t_0, t_1]$.

In link prediction, $ET$ is treated as known information while $EP$
is only used to test the accuracy. For each pair of nodes $u$, $v$ $
\in V'$ without a link in $G[t_0, t_1]$, each link predictor that we
consider assigns a similarity score based on the existing links in
$ET$. Then all unlinked pairs are ranked in descending order
according to their scores. In this study, we use two evaluation
metrics, AUC (Area Under the Receiver operating characteristic
curve) and precision. In the present case, AUC can be simplified as
the probability that a randomly chosen link in $EP$ has higher
similarity score than a randomly chosen nonexistent link in $
{U\backslash (ET \cup EP)} $. In the evaluation implementation,
among $n$ times of independent comparisons, if there are $n'$ times
that the future link has a higher score and $n''$ times the future
link and nonexistent link have the same score, then AUC can be
calculated by ${\rm{AUC}} = \frac{{n' + 0.5n''}}{n}$. If all the
scores are generated from an independent and identical distribution,
AUC will be approximately 0.5. Therefore, the extent to which AUC
exceeds 0.5 indicates how much better the algorithm performs than
pure chance. We compare the first $L$, $L = |EP|$, links with $EP$
to compute the precision of a predictor. Given the ranking of the
non-observed links ${U\backslash ET}$ in which there are some future
links, if $Lr$ links among top-$L$ links are accurately predicted
($Lr$ is the number of the top-$L$ links in the probe set $EP$),
then ${\rm{Precision}} = Lr/L$. In this paper, our fundamental
hypothesis is that the addition of a set of links does not
significantly change network's structure features, and we make sure
training set representing a connected network in empirical analysis.

\subsection{Similarity indices}\indent

Among many similarity indices, Liben-Nowell and Kleinberg [15] and
Zhou et al [19] showed that Common Neighbors (CN), Adamic-Adar (AA)
and Resource Allocation (RA) indices perform best by systematically
comparing local similarity indices in unweighted networks.
Therefore, in this paper, we concentrate on the weighted definition
of these three indices (denoted by WCN, WAA, WRA respectively) and
more details can be found in Ref. [12].

\begin{equation}
s_{xy}^{WCN}  = \sum\limits_{z \in \Gamma (x) \cap \Gamma (y)}
{w(x,z) + w(z,y)}
\end{equation}
\begin{equation}
s_{xy}^{WAA}  = \sum\limits_{z \in \Gamma (x) \cap \Gamma (y)}
{\frac{{w(x,z) + w(z,y)}}{{\log (1 + s(z))}}}
\end{equation}
\begin{equation}
s_{xy}^{WRA}  = \sum\limits_{z \in \Gamma (x) \cap \Gamma (y)}
{\frac{{w(x,z) + w(z,y)}}{{s(z)}}}
\end{equation}

\noindent Here, ${w(x,y) = w(y,x)} $ denotes the weight of the link
between node $x$ and $y$, and $s(z) = \sum\nolimits_{z' \in \Gamma
(z)} {w(z,z')}$ denotes the strength of node $z$, namely the sum of
weights of its associated links. Moreover, we also take Local Path
(LP) [19] index into account:

\begin{equation}
s_{xy}^{LP}  = A^2  + \varepsilon  \cdot A^3
\end{equation}

\noindent where $A$ is the adjacent matrix of network, and $e$ is a
free parameter. LP index makes use of the information on local paths
with lengths 2 and 3. In the real implementation, we directly count
the number of different paths with length 2 and 3 and the parameter
is fixed at $\varepsilon={\rm{10}}^{ - 3}$ following the original
article [19]. In weighted networks, we sum up the weights of
different paths with length 2 and 3.

\section{Method}\indent

Prediction of evolving networks attempts to predict future network
structure by mining data on past node interactions. From the
perspective of dynamics, the dynamical evolution of complex networks
can be simply given by the dynamical states in the phase space of
the network, and the network dynamics can be decomposed into
products of single node dynamics. What most have in common of the
existing indices is that they fix the node similarities in
prediction process. Here, we present a novel angle of view for
future links prediction. The basic philosophy is to envision
evolving networks as dynamic systems and dynamically investigate
nodes' similarities to uncover network's evolutionary trend. The
viewpoint provides a vivid and intuitive image to model the
real-world network evolution. Before introducing the dynamics
evolutionary model, we firstly present a rather robust and general
similarity index by exploiting different range of structure
information under the guidance of network macroscopical properties
as follows.

\subsection{Structure-dependent similarity index}

Although for many networks with high clustering coefficient,
neighbor-based methods, such as CN, AA, RA, can obtain satisfactory
prediction accuracy. However, in some sparse networks with low
clustering coefficient, it is difficult for neighbor-based methods
to achieve high prediction accuracy. This may be because such
neighbor-based indices cannot calculate similarity between nodes
without common neighbors. Moreover, the neighbor-based methods are
less distinguishable from each other, and the probability that two
node pairs are assigned the same score usually is high [13]. To
resolve the weaknesses of lower prediction accuracy, some
distance-based methods are proposed, such as LP [19], Katz [20] and
LHN-II [17]. However, the distance-based similarity indices are
always sensitive to the proportion of observed edges. It means that
their prediction accuracies will reduce obviously if the proportion
of observed links decreases in training set. This is because the
removal of links will increase the average shortest distance between
node pairs. Thus, if the path range of distance-based indices
smaller than the shortest distance between node pairs, the indices
entirely cannot capture the similarities between the node pairs.

Based on the above analysis, it is necessary that proposing a rather
robust and general similarity index for future links prediction. To
strengthen the robustness of indices and calculate the similarity of
node pairs in networks with different structure properties,
similarity indices should count all possible paths between the
nodes. However, the computation complexity of similarity index
increases exponentially with the growth of the range of paths. As a
result, we propose a structure-dependent (SD) similarity index
exploiting different range of local structure information based on
network macroscopical properties. The formal definition is given as
follows:

\begin{equation}
s_{xy}^{SD}  = \frac{{A^s }}{{\overline {k(z)} }} + \varepsilon
\cdot \frac{{A^{s + 1} }}{{\overline {k(z)} }}
\end{equation}

\noindent where $A$ is the adjacent matrix, $k(z)$ denotes the
degree of node $z$, $\overline {k(z)}$ is the average degree of the
intermediate nodes between $x$ and $y$, $s$ is the shortest distant
between node $x$ and $y$ and $s \le \max(<d>,2)$, $<d>$ is the
average shortest distant between node pairs. The weighted
structure-dependent (WSD) index is defined as follows:

\begin{equation}
s_{xy}^{WSD}  = \frac{{W^s }}{{\overline {s(z)} }} + \varepsilon
\cdot \frac{{W^{s + 1} }}{{\overline {s(z)} }}
\end{equation}

\noindent where $W$ is the weight matrix, $s(z)$ denotes the
strength of node $z$, $\overline {s(z)}$ is the average node
strength of the intermediate nodes between $x$ and $y$,
$\varepsilon$ is a free parameter. In the real implementation, we
directly sum up the weights of the different paths with length $s$
and $s+1$ and the parameter is fixed at $\varepsilon={\rm{10}}^{ -
3}$ following the definition of LP [19]. According to the
definition, we can find that WSD captures the shortest paths and the
next shortest paths for similarity calculation. Furthermore, WSD
reduces to WLP when the shortest distant is two and the average node
strength $\overline {s(z)}$ equals to one, and WSD index reduces to
WRA when range of the shortest distant is two and the parameter
$\varepsilon$ is zero.

\subsection{Spatial-temporal network position drift model}

Spatial-temporal network position drift model is about the problem
of how network nodes displace their network position to represent
the potential trend of network evolution considering the spatial and
temporal factors. Nodes' network position is the similarity
relationships between the nodes and their neighbors. As nodes
interact only with their neighbors, the migration and variation of
nodes' network position can only be achieved via local interactions.
Moreover, assuming that the link weight of two connected nodes
indicates their interaction strength and the larger link weight
represents the closer relationship between them, we set the link
weights as the initial similarities of the connected nodes.

\begin{figure}[!h]
\centering \subfigure[]{ \label{fig:side:a}
\includegraphics[width=2.2in]{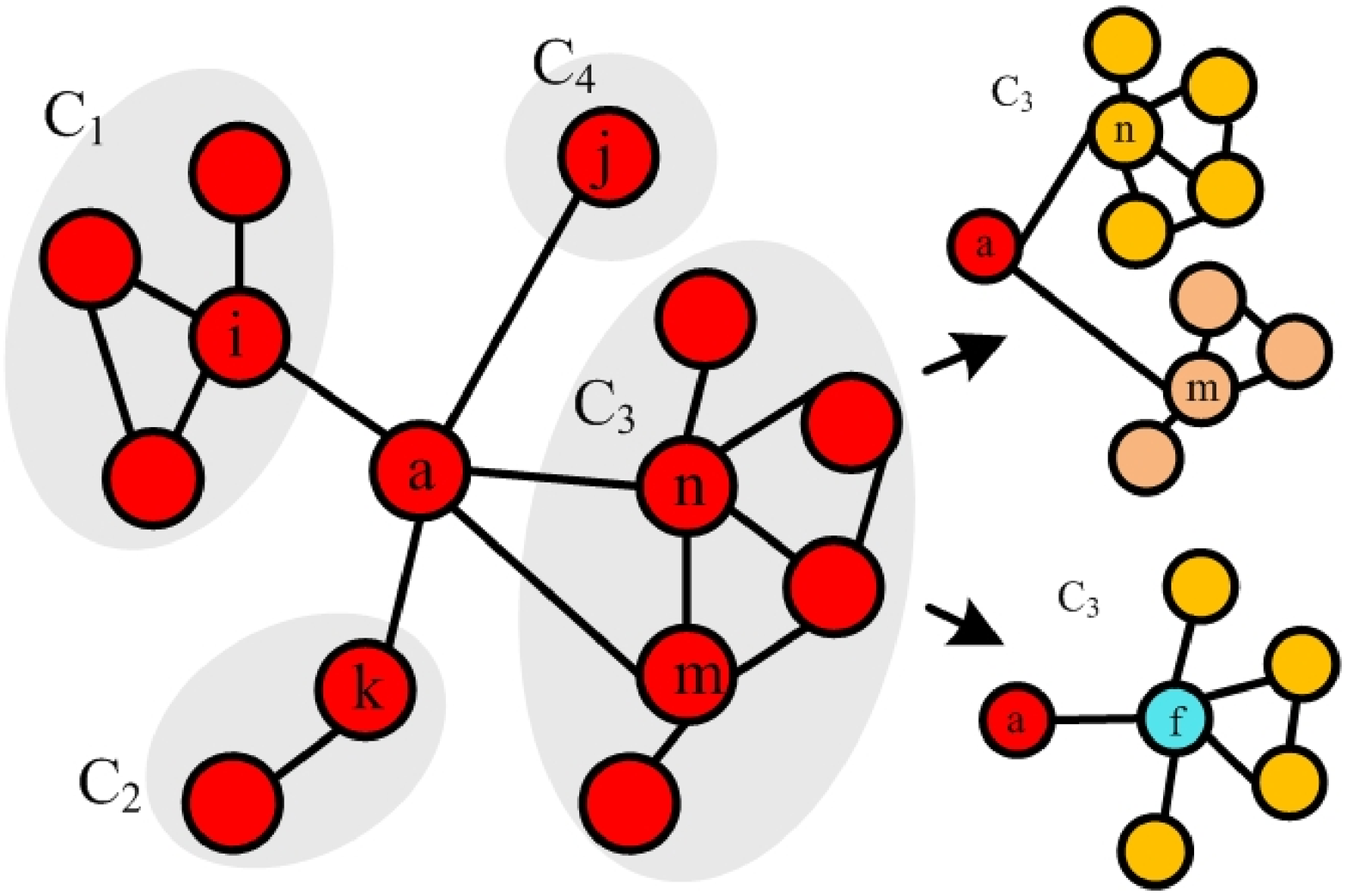}}
\hspace{1ex} \subfigure[]{ \label{fig:side:b}
\includegraphics[width=2.1in]{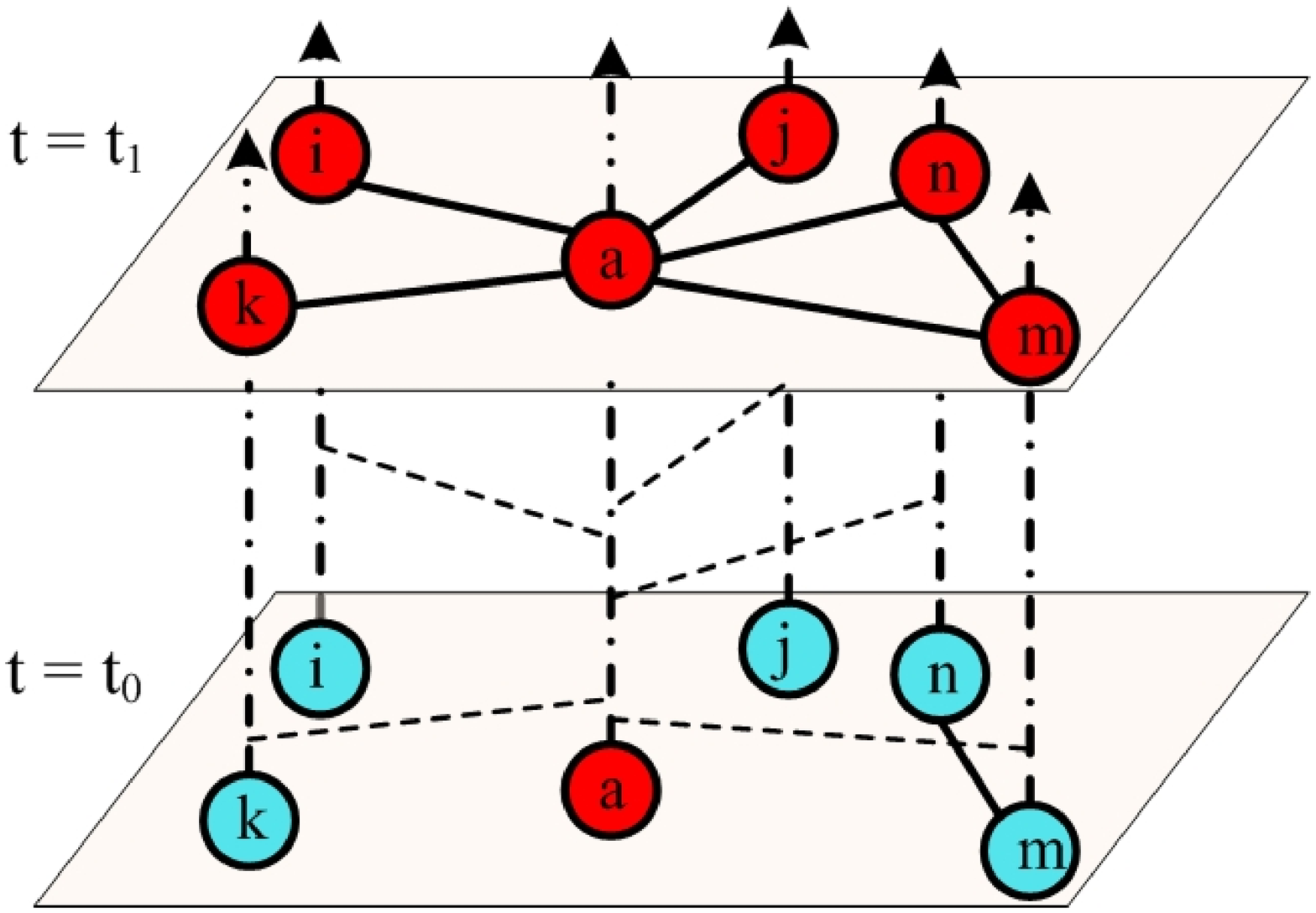}}
\hspace{1ex} \subfigure[]{ \label{fig:side:b}
\includegraphics[width=2.2in]{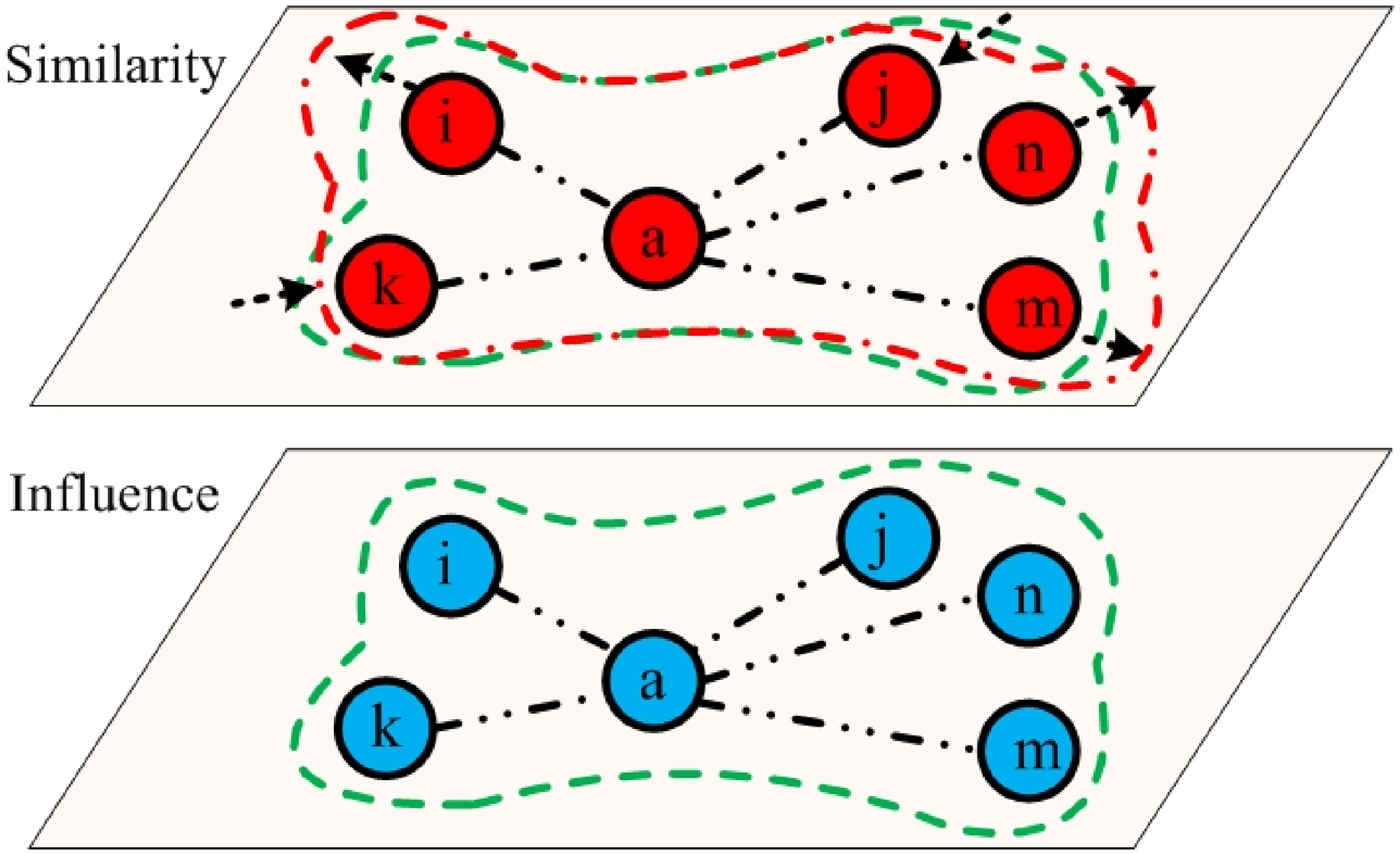}}
\caption{The illustration of the spatial-temporal interaction model.
(a) The graph representation of spatial interaction within local
topology. (b) The graph representation of temporal interaction. (c)
Force difference based spatial-temporal network position drift.
(Arrows indicate the force difference between spatial-temporal
influence and bond strength, and the node distance in the two
subgraphs is proportional to the similarity and influence strength
between them respectively.)}\label{fig:side} \vspace{\baselineskip}
\end{figure}

\textbf{Spatial factor.} To model the spatial influence, this paper
assumes that network nodes are placed in force fields. To a network
node, the attractiveness of all the node's neighbors influence the
central node's network position. According to Ref. [27, 28], network
nodes tend to approach the close connected groups and be far away
from the uncompacted ones. Based on in-depth analysis of local
structure, we find that each node has different network connection
ability for structure organization. Commonly, the higher clustering
level of groups that a node belongs to, the more and the stronger
interactions within the node' neighborhood, and the higher
connection ability of the node. So network nodes tend to approach
the nodes with high connection ability from the perspective of
microstructure. To measure the influence of network nodes, this
paper calculates the connection abilities of network nodes as their
attractiveness. As nodes interact only with their neighbors, every
network node takes all the attractiveness of its neighbors into
account and then decide the direction and scale of position drift.
Let's take central node $a$'s neighbor node $i$ and $j$ in Fig. 1(a)
for example. There are multiple edges in the one-hop neighborhood of
node $i$, and $j$ is an isolated node. Thus node $i$ has higher
connection ability and attractiveness than that of node $j$, and
node $a$ will tend to approach node $i$ rather than node $j$. It is
also possible that the interactions between node $a$ and the
neighbors of node $i$ will be encouraged and strengthened. Formally,
considering the link number and the  link weight in one-hop
neighborhood of network nodes, the attractiveness of them is defined
as follows:

\begin{equation}
AI(j)  = cn(j) \cdot st(j) = \sum\limits_{l \in Edge(ego(j))} {w_l }
\end{equation}

\noindent where $ego(j)$ is the set of nodes in the one-hop
neighborhood of node $j$, $Edge(ego(j))$ is the set of links between
the nodes in $ego(j)$, $w_l$ is the weight of link $l$, $cn(j) =
|E(ego(j))|$ is the connectivity ability of node $j$, and $ st(j) =
\sum\limits_{l \in E(ego(j))} \frac{{w_l}}{|E(ego(j))|} $ is the
connectivity strength of node $j$.

Except for the interactions between the central node and its
neighbors, there also may have interactions between the neighbors.
In such situation, the neighbors and the central node form a triadic
closure, and the connected neighbors would influence the central
node consistently. The integrated attractiveness of the connected
neighbors to the central node is different from the independent
influence of anyone of the neighbors. Let's take the neighbor node
$m$ and $n$ in Fig. 1(a) for example, here, node $m$ and $n$ have a
consistent and common attractiveness to the central node $a$ in Fig.
1(a), which is different from the independent influence of node $m$
or $n$ shown in the top right corner of Fig. 1(a).

To measure the attractiveness of the node $m$ and $n$ to the central
node $a$, node $m$ and $n$ is considered as a whole to analyze. We
fuse the neighborhood of node $m$ and $n$ as a new node $f$, as
shown in the bottom right corner of Fig. 1(a), and measure the
attractiveness of the new node to denote the attractiveness of node
$m$ and $n$ proportionally. Formally, the attractiveness of a node
in the connected neighbor set can be calculated by:

\begin{equation}
AI(i) = |NC| \cdot AI(v) \cdot \frac{{AI'(i)}}{{\sum\nolimits_{j \in
NC} {AI'(j)} }}
\end{equation}

\noindent where $NC$ is the connected neighbor set, ${AI(v)}$ is the
attractiveness of the virtual node fused from $NC$, and ${AI'(i)}$
is the attractiveness of node $i$ in independent situation. Let's
take node $m$ and $n$ in Fig. 1(a) for example, $ AI(m) = 2AI(f)
\cdot \frac{{AI'(m)}}{{AI'(m) + AI'(n)}} $, $ AI(n) = 2AI(f) \cdot
\frac{{AI'(n)}}{{AI'(m) + AI'(n)}} $, where ${AI(f)}$ is the
attractiveness of the new node, $AI(f) = \sum\limits_{l \in
Edge(ego(m)) \cup Edge(ego(n))} {w_l } $, ${AI'(m)}$ and ${AI'(n)}$
are the attractiveness of node $m$ and $n$ in independent situation
as shown in the top right corner of Fig. 1(a). Clearly, we have
$AI(m) + AI(n) = 2AI(f)$. What is worth noticing is that the
connected neighbors have common intention and consistent behavior in
attracting the central node.

\textbf{Temporal factor.} According to the results of Ref. [29, 30],
this paper assumes that the evolution process of networks is
continuous and future state of evolving network is more relevant
with the current state than the past states in evolution trend
prediction. That is to say, more recent interactions are more
powerful for potential trend indication of evolving networks than
the older ones. To the central node $a$ in Fig. 1(a), the temporal
dynamics can be illustrated by Fig. 1(b), and the interaction set of
node $a$ can be represented as \{$l_{aj}$, $l_{ai}$, $l_{an}$,
$l_{ak}$, $l_{am}$\} in timestamp sequence. With the growth of
survival time, the indicative power of the interactions is reduced
gradually. As a result, the temporal importance of link $l_{ai}$ for
future links prediction is defined as

\begin{equation}
PI(l_{ai} ) = \frac{{e^{(t(l_{ai} ) - \overline {t(\{ l_a \} )}
)/2\Delta t} }}{{1 + e^{(t(l_{ai} ) - \overline {t(\{ l_a \} )}
)/2\Delta t} }}
\end{equation}

\noindent where $\overline {t(\{ l_a \} )}$ is the average of
timestamps $\{ l_a \}$, $\overline {t(\{ l_a \} )} =
\frac{{\sum\nolimits_{j \in N(a)} {t(l_{aj} )} }}{{|N(a)|}}$,
$\Delta t$ is the unit time interval, $\Delta t = \frac{{\mathop
{Max}\limits_{i \in N(a)} t(l_{ai} ) - \mathop {Min}\limits_{j \in
N(a)} t(l_{aj} )}}{{|N(a)|}}$, $t(l_{ai} )$ is the timestamp of
interaction between node $a$ and $i$, and $N(a)$ is the neighbor set
of node $a$.

Based on the above analysis, the spatial-temporal influence of each
neighbor $j$ on central node $i$ is comprehensively defined as:
\begin{equation}
A(j) = \frac{{AI (j)}}{{\mathop {\max }\limits_{k \in N(i)} AI (k)}}
\cdot \frac{{PI(l_{ij})}}{{\mathop {\max }\limits_{k \in N(i)} PI
(l_{ik})}}
\end{equation}

\textbf{Spatial-temporal Position Drift.} After modeling the
spatial-temporal influence of every neighbors, how should nodes
drift their network position according to the spatial-temporal
influence? As analyzed in the local interaction process, all
neighbors have spatial-temporal influence on the central node. In
fact, every neighbor attracts the central node to move towards
itself through mutual interactions. In order to establish the
optimum trade-off among the spatial-temporal influence of every
neighbors, the neighborhood of the central node can be regarded as a
force field of node influence. Meanwhile, the similarities between
the central node and its neighbors denote the current bond strength
of them. Motivated by comparison of the spatial-temporal influence
and the bond strength in the neighborhood of the central node, we
argue that the direction and scale of position drift are all arise
from the force difference between them. If the influence of an
neighbor node on the central node is greater than the bond strength
between them, the similarity between them increases. Conversely, the
similarity decreases. As illustrated by Fig. 1(c), the arrows
indicate the force difference between the spatial-temporal influence
and the bond strength, where the distance between node pairs is
proportional to the similarity and influence strength between them.
Based on the above philosophy, we define $\Delta {{s(a, j)}}$ to
characterize the dynamic of similarity ${{s(a, j)}}$ between node
$a$ and $j$, ${{j}} \in {{N(a)}}$:
\begin{equation}
\Delta s(a, j)  =  - {\sum\limits_{k \in N(a)} {s(a, k) } }
(\frac{{s(a, j) }}{{\sum\limits_{k \in N(a)} {s(a, k) } }} -
\frac{{A(j) }}{{\sum\limits_{k \in N(a)} {A(k) } }})
\end{equation}

\section{Empirical Analysis}\indent
In this section, to demonstrate the benefits of the proposed method,
we apply it on real-world networks for empirical analysis.

\subsection{Data description}

The dataset studied in this paper, including two static networks and
three evolving networks, are detailed as follows. (1) Celegans [8]:
The neural network of C. Elegans. (2) USAir [31]: The US Air
transportation network. (3)MIT [32]: The network contains human
contact data among 100 students of the Massachusetts Institute of
Technology (MIT), collected by the Reality Mining experiment
performed in 2004 as part of the Reality Commons project. (4)
Hypertext 2009 network [33]: The network represents the face-to-face
contacts of the attendees of the ACM Hypertext 2009 conference. Node
represents a conference visitor, and an edge represents a
face-to-face contact that was active for at least 20 seconds. Each
edge is annotated with the time at which the contact took place. (5)
Infectious network [33]: The network describes the face-to-face
behavior of people during the exhibition INFECTIOUS: STAY AWAY in
2009 at the Science Gallery in Dublin. Nodes represent exhibition
visitors, and edges represent face-to-face contacts that were active
for at least 20 seconds. The basic structure properties of the
networks are summarized in Table 1. The original networks are turned
into undirected and simple networks.

\begin{table}[!h]\small \centering
\caption{The basic structure properties of the giant components of
the five example networks. $N$ and $M$ are the total numbers of
nodes and links, respectively. $\langle k\rangle$ is the average
degree of the networks. $\langle d\rangle$ is the average shortest
distant between node pairs. $C$, $C_w$ and $r$ are clustering
coefficient, weighted clustering coefficient and assortative
coefficient, respectively. $H$ is the degree heterogeneity, defined
as $H = \frac{{\langle k^2 \rangle }}{{\langle k\rangle ^2 }}$,
where $\langle k\rangle$ denotes the average degree.}
\includegraphics[width=5.5in]{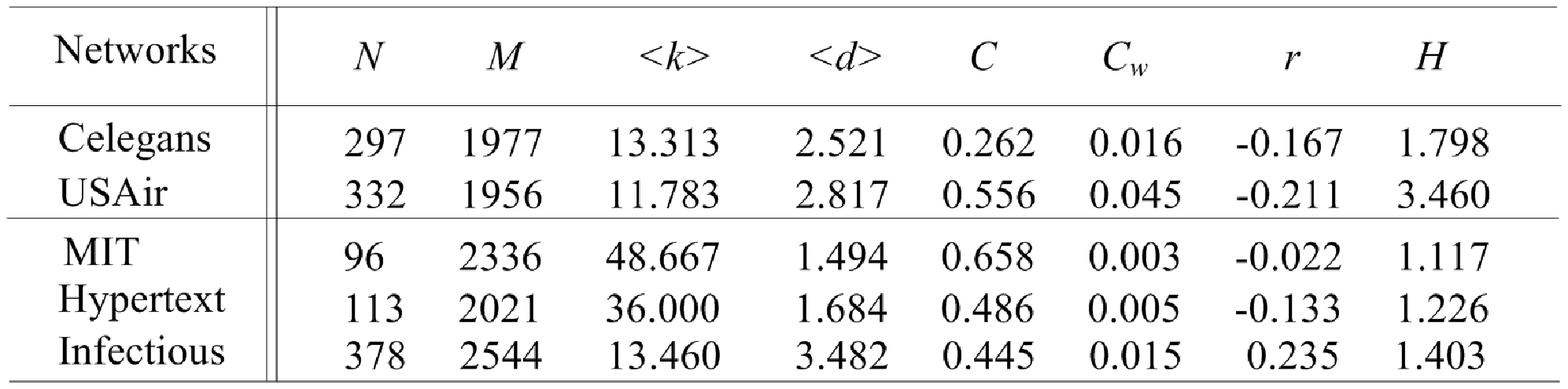}
\end{table}

\subsection{Effectiveness and robustness of similarity index WSD}

To verify the effectiveness of the proposed index WSD, we compare
the prediction accuracy of different similarity index under the AUC
metric and the Precision metric. The results are shown in Table 2
and Table 3 respectively. The metrics are introduced in the
Preliminaries section. The highest AUC/Precision value for each
network is shown in boldface. Under the AUC metric, WSD performs
best in 3 out of 5 networks and performs next best in the other 2
networks, while WSD performs best in all of them under the Precision
metric. In addition, we compare the prediction accuracy of different
similarity index under varied ratio of deleted links in training
set, and Fig. 2 shows the results of the methods. It can be seen
that the proposed index is either the best or very close to the best
in the five real-world networks. So we can conclude that the
proposed WSD index is competitive to the state-of-the-art methods.

\begin{table}[!h]\small \centering
\caption{Comparison of the prediction accuracy under the AUC metric
in real-world networks.}
\includegraphics[width=6.3in]{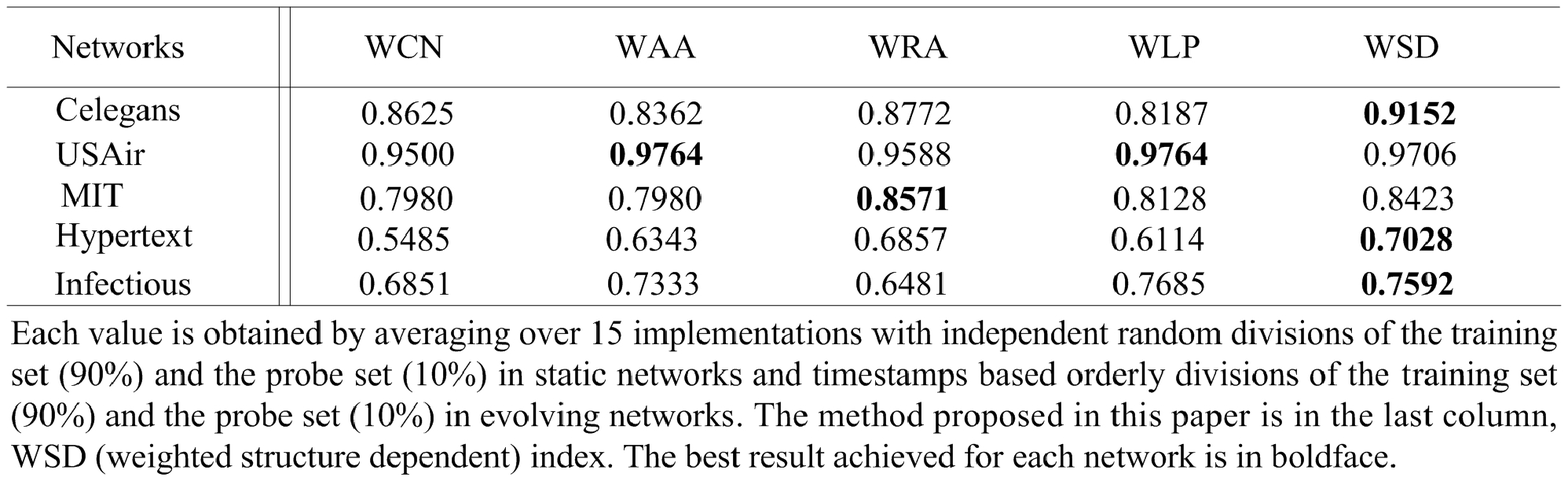}
\end{table}

\begin{table}[!h]\small \centering
\caption{Comparison of the prediction accuracy under the Precision
metric in real-world networks.}
\includegraphics[width=6.3in]{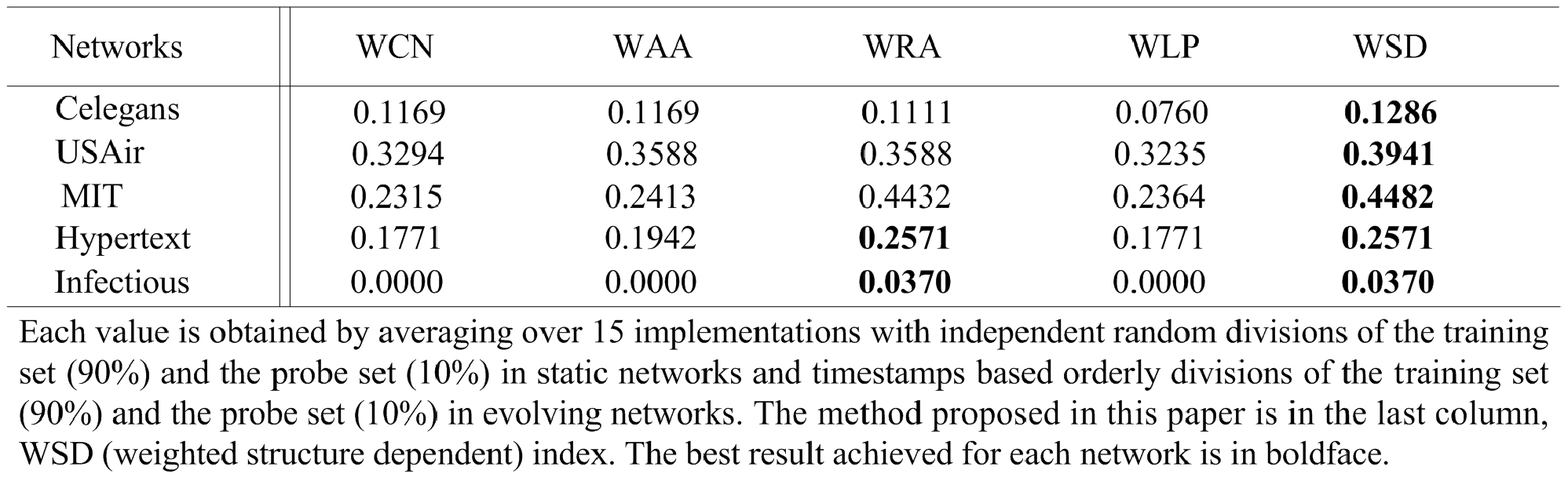}
\end{table}

\begin{figure}[!h]   \small \centering
\includegraphics[width=7.2in]{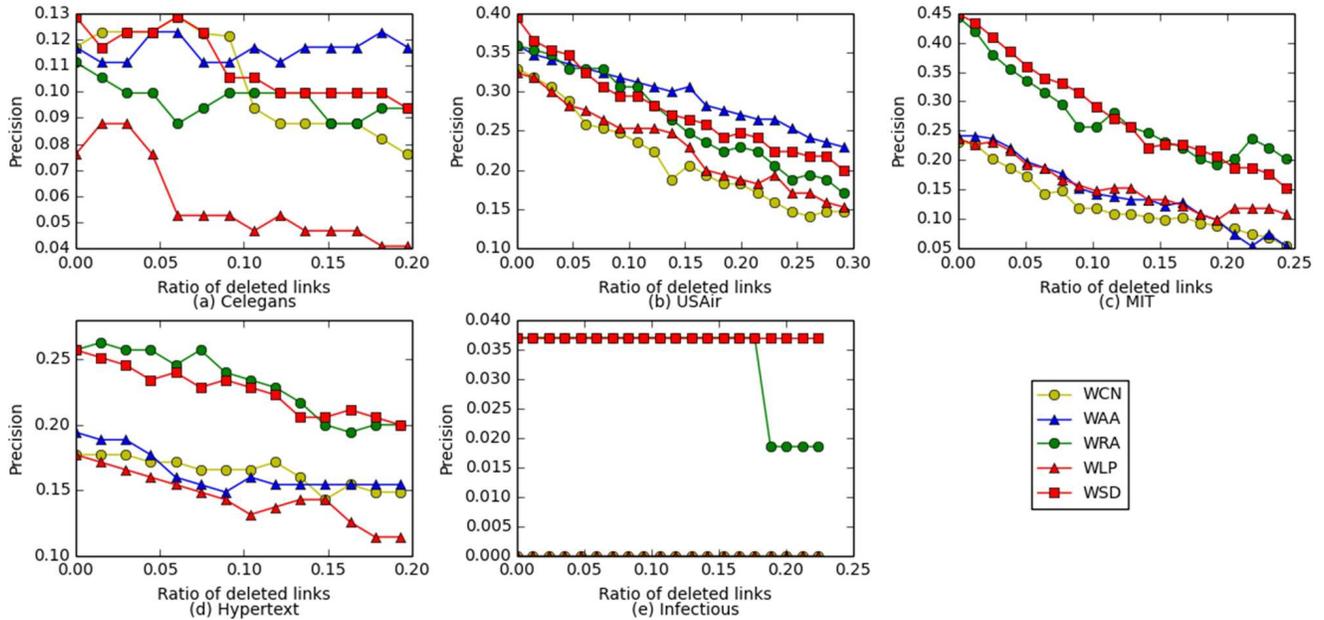}
\caption{Effectiveness of similarity index WSD under varied ratio of
deleted links in five real-world networks. (The horizontal axis
denotes the ratio of deleted links in training set for future links
prediction. Each value of the Precision is a result averaged over 15
implementations with independent random divisions of the training
set (90\%) and the probe set (10\%) in static networks and
timestamps based orderly divisions of the training set (90\%) and
the probe set (10\%) in evolving networks.)}\label{fig:2}
\end{figure}

\begin{figure}[!h]
\centering \subfigure[Robustness of similarity index WSD with
different ratio of deleted links]{ \label{fig:side:a}
\includegraphics[width=7.1in]{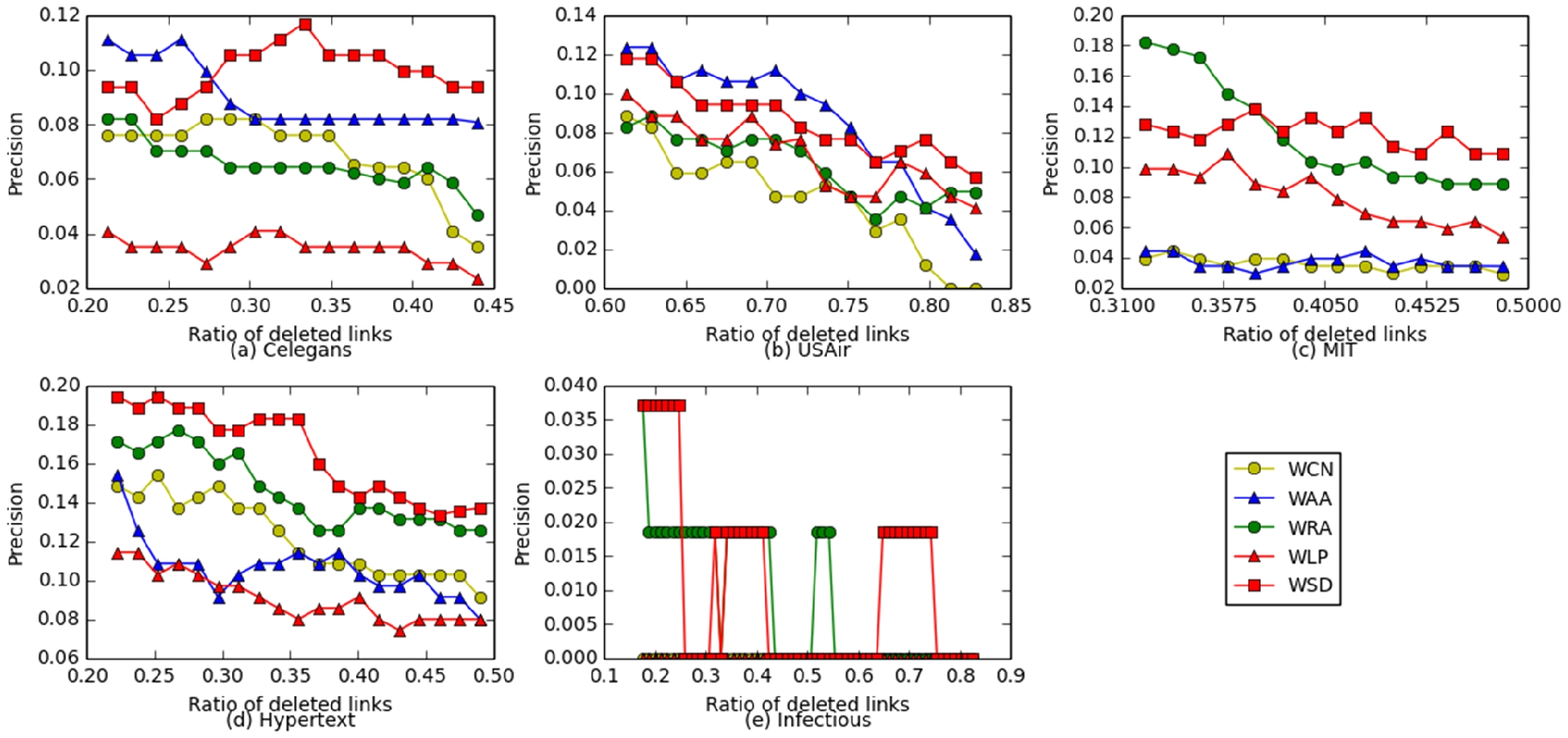}}
\hspace{1ex} \subfigure[Average shortest distant between node pairs
with different ratio of deleted links]{ \label{fig:side:b}
\includegraphics[width=7.1in, height=3.1in]{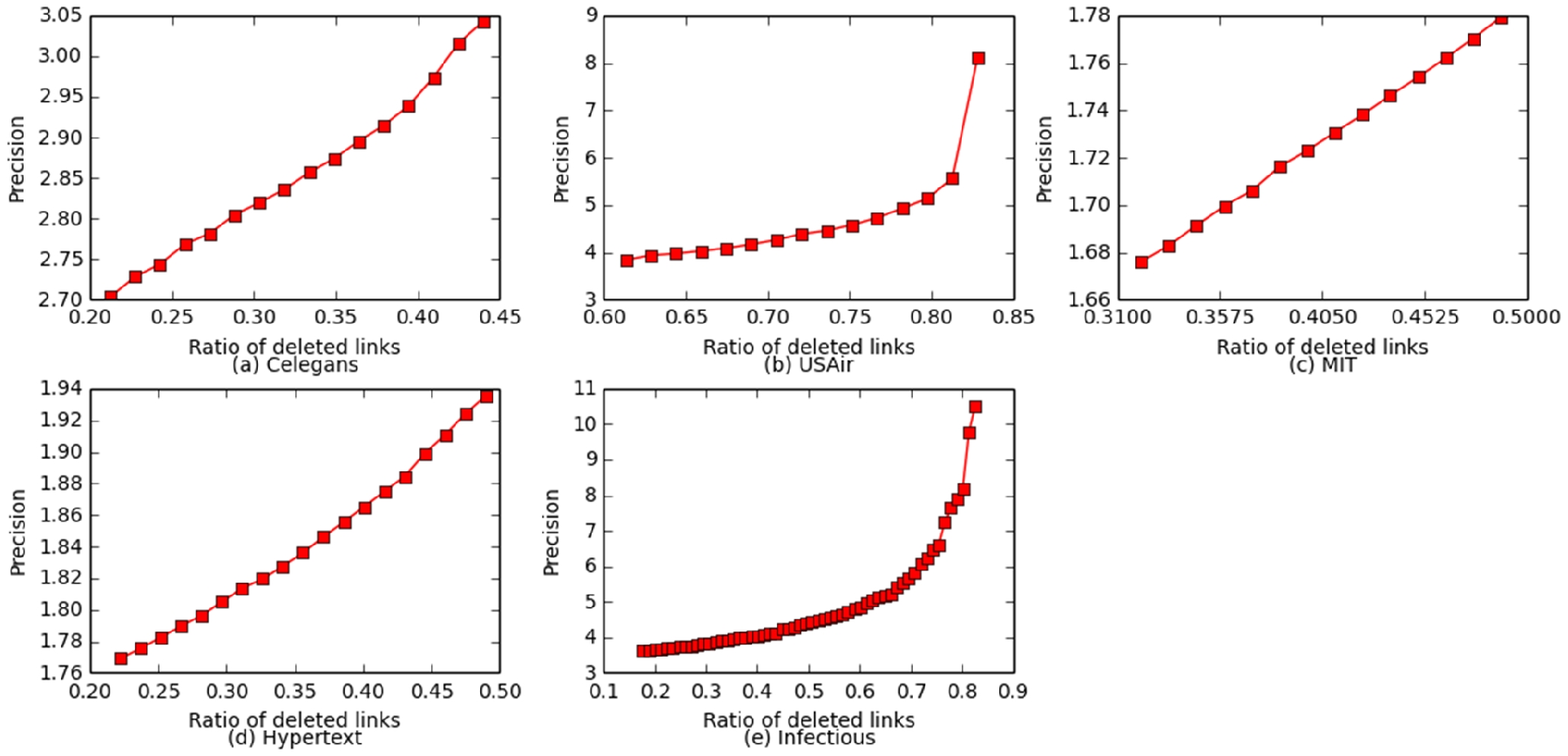}}
\caption{The illustration of the robustness of similarity index WSD.
(The horizontal axis denotes the ratio of deleted links in training
set for future links prediction. Each value of the Precision is a
result averaged over 15 implementations with independent random
divisions of the training set (90\%) and the probe set (10\%) in
static networks and timestamps based orderly divisions of the
training set (90\%) and the probe set (10\%) in evolving
networks.)}\label{fig:side} \vspace{\baselineskip}
\end{figure}

As discussed in section 3.1, neighbor-based indices cannot calculate
the similarity of nodes without common neighbors and distance-based
indices cannot capture the similarity of nodes if the path range of
the distance-based indices is smaller than the shortest distance
between the nodes. For example, looking at Table 3 and Fig. 2(e),
for the prediction precision of the links in network Infectious,
WCN, WAA and WLP are completely useless and the performance of WRA
is also unsatisfactory. The main reason is that the average shortest
distant 3.482 of network Infectious is greater than the path range
2.0 of similarity indices WCN, WAA, WRA and the the path range 3.0
of similarity index WLP. Hence, the indices cannot capture the
similarities of the nodes in network Infectious. In contrast, WSD
captures the shortest paths and the next shortest paths for
similarity calculation under the guidance of networks' structure
properties.

Fig. 3 inspects the robustness of the proposed index WSD by
comparing the prediction accuracy of different indices under varied
ratio of deleted links. Intuitively, the more the amount of the
deleted links, the less the amount of known information and the
larger the average shortest distant. Fig. 3(a) describes a general
trend of descending in prediction precision and Fig. 3(b) describes
a general trend of ascending in average shortest distant with the
growth of the ratio of deleted links. Synthesizes the results of
Fig. 3(a) and Fig. 3(b), what calls for special attention is that
WSD has satisfactory performance in the real-world networks and its
advantage comparing the other similarity indices increase gradually
with the increase of the average shortest distant. So WSD is robust
in real-world networks with different structure properties.

\subsection{The effectiveness of position drift on future links prediction}\indent

\begin{figure}[!h]   \small \centering
\includegraphics[width=6.2in]{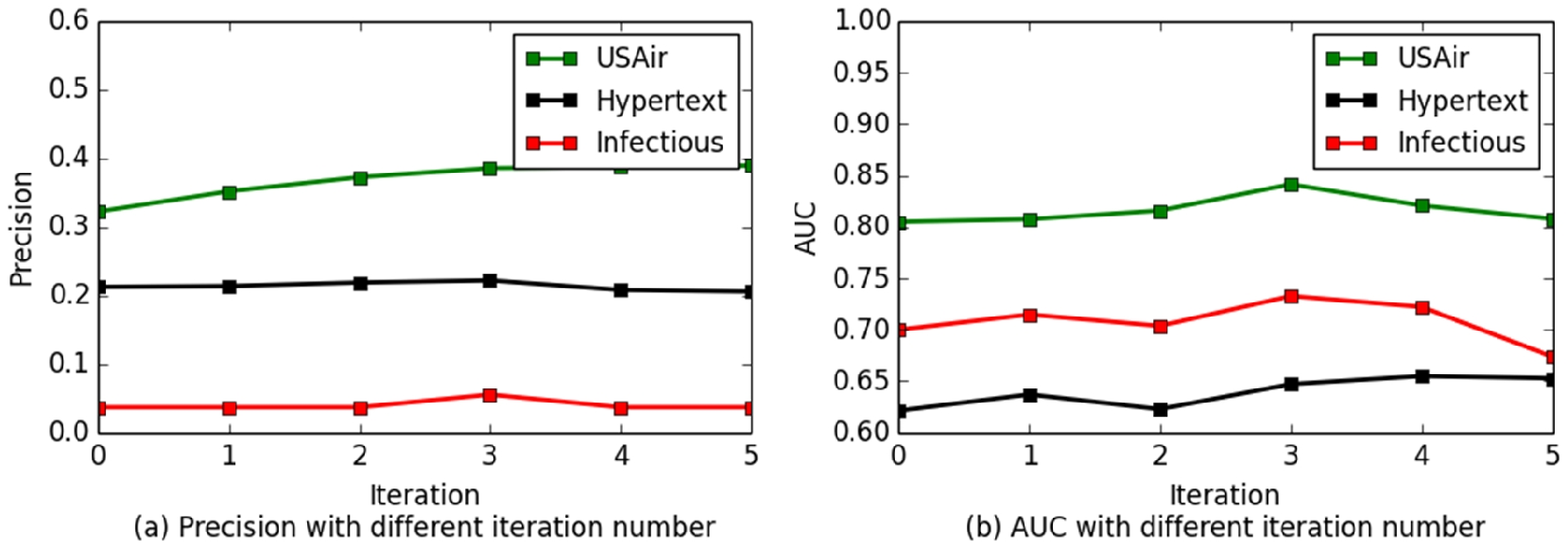}
\caption{The sensitivity of iteration number of position drift on
link prediction. (Each value of the Precision and AUC is a result
averaged over the similarity indices and 15 independent
implementations with timestamps based orderly divisions of the
training set (90\%) and the probe set (10\%) in evolving
networks.)}\label{fig:4}
\end{figure}

For uncovering the potential trend of network evolving, network
nodes update their network position iteratively according to
spatial-temporal position drift model in Eq. (11). Then link
prediction methods calculate node similarity based on the drifted
network topology. As  different iteration number of position drift
would lead to different network topology and different prediction
accuracy, here, we study the influence of the iteration number on
the evolving network prediction. The result is plotted in Fig. 4
with iteration number ranging from 0 to 5 on three evolving
real-world networks. From this plot, we can see that the overall
optimal performance is got when the iteration number equals to 3.

To each evolving network, every network node apply the position
drift model to displace their network position firstly, then future
links are predicted using the similarity indices on the drifted
network. We compare the prediction accuracy with the result obtained
based on the original network topology under the AUC metric and the
Precision metric. From Table 4 and Fig. 5, it can be seen that the
prediction results based on the drifted networks are either close to
or better than the result obtained based on the original network
topology, and the spatial-temporal position drift is effective for
the evolving networks.

\begin{table}[!h]\small \centering
\caption{The effectiveness of position drift on future links
prediction under the AUC metric.}
\includegraphics[width=6.6in]{tb3.eps}
\end{table}

\subsection{Running time}\indent

Table 5 presents the computation time of the link prediction methods
and the position drift model with varied iteration number. As WAA,
WRA are variants of WCN, they have nearly the smallest computation
values in the real-world networks. WLP captures the information of
paths with two and three hops and needs more computation time than
that of WCN, WAA, WRA. Compared with WLP, the WSD index achieves
competitive performance in all the networks. In addition, to
position drift model, each node displace its network position based
on the attractiveness of its neighbors, so its computational
complexity is sensitive to the average degree of network. According
to Table 5, we can find that the computation times on the three
evolving networks decrease with the decline of the average degree of
the networks. Since the real-world networks are mostly sparse, the
position drift model is practical in realistic application. In
summary, the proposed prediction index and the position drift model
are all practical from the perspective of computation time.

\begin{figure}[!h]   \small \centering
\includegraphics[width=7.0in]{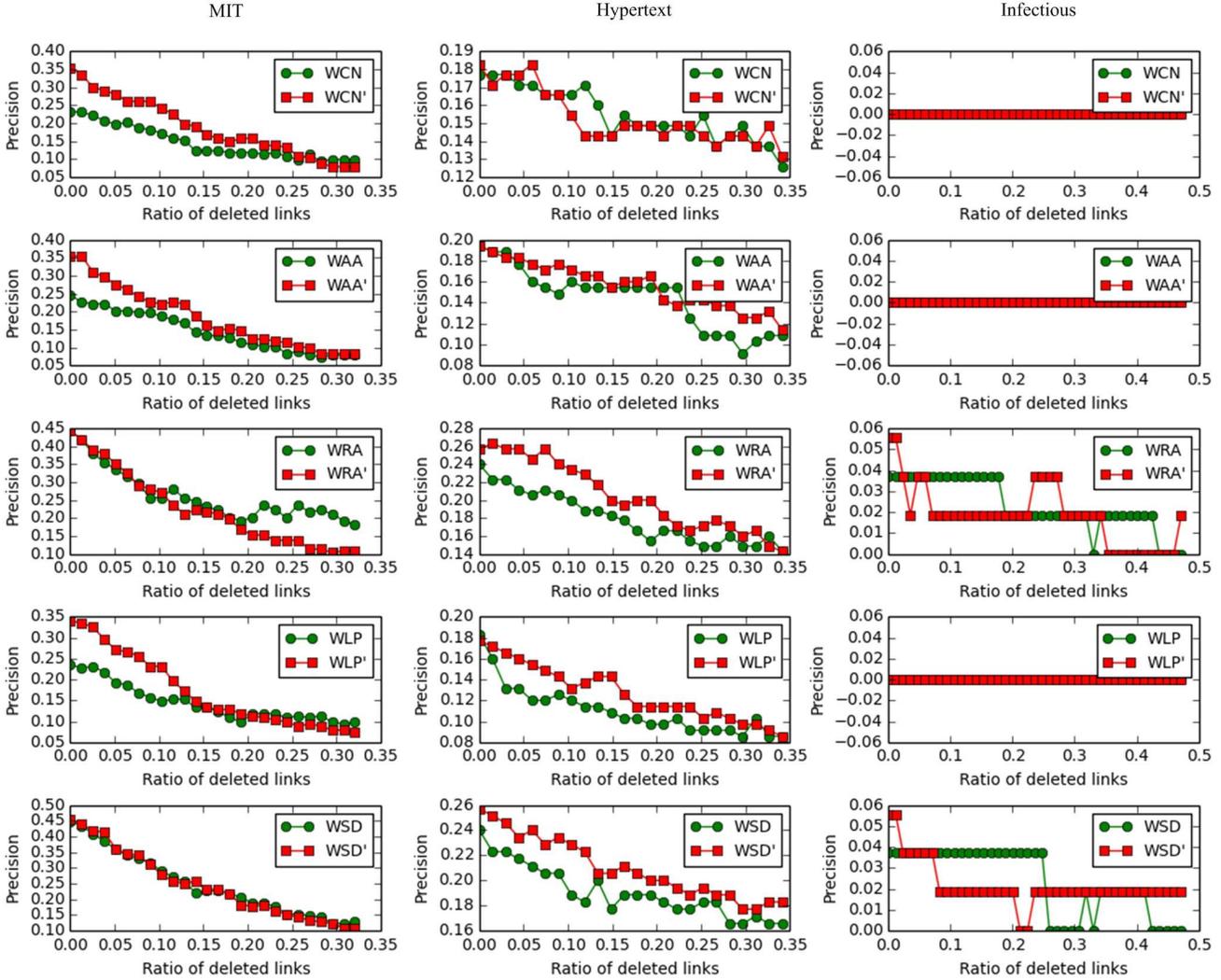}
\caption{The effectiveness of position drift on future links
prediction under the Precision metric. (The horizontal axis denotes
the ratio of deleted links in training set for future links
prediction. Each value of the Precision is a result averaged over 15
independent implementations with timestamps based orderly divisions
of the training set (90\%) and the probe set (10\%) in evolving
networks. The iteration number of position drift is fixed as 3.
WCN', WAA', WRA', WLP', WSD' denotes the prediction manipulation of
WCN, WAA, WRA, WLP, WSD on drifted networks.)}\label{fig:4}
\end{figure}

\begin{table}[!h]\small \centering
\caption{Computation time (in millisecond) comparison on the
real-world networks.}
\includegraphics[width=5.4in]{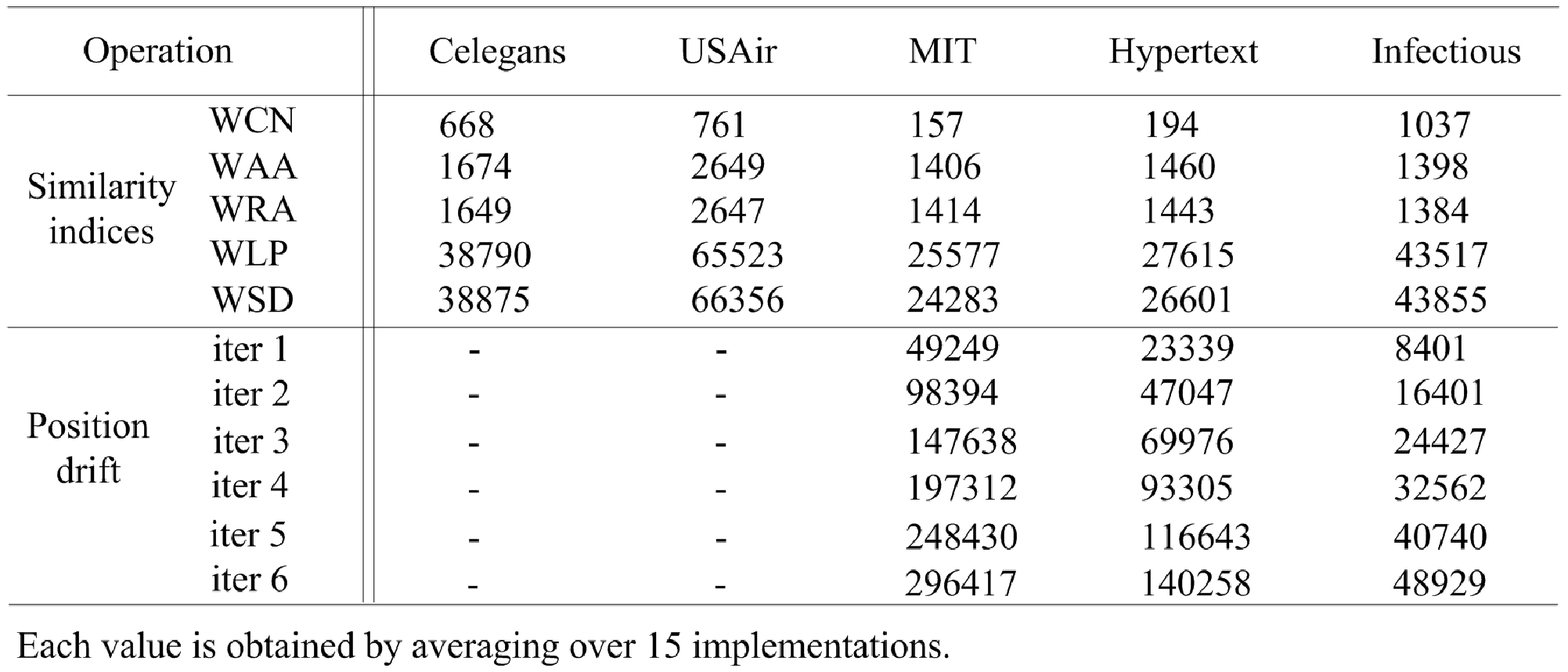}
\end{table}

\section{Discussion and Conclusion}\indent

Evolution of networks are widely studied to effectively understand
and predict the potential trend of complex systems. Although many
link prediction algorithms have been proposed in the literature, the
prediction of evolving networks still remains unclear. To address
this issue, the paper firstly proposed a robust similarity index for
ranking candidate links in different networks. Then the paper views
the evolving networks as dynamic systems and develops a network
position drift model to uncover the potential trend of evolving
networks. Experimental results show that the proposed methods
perform well in real-world networks.

The model developed in this paper, though simple, captures several
theoretically important features of prediction in evolving networks,
such as the clustering effect and the novelty effect of network
interactions. It is worth noticing that the proposed position drift
model does not adopt clustering coefficient to measure the
attractiveness of neighbor nodes, because the clustering coefficient
values of the nodes which do not have closed triadic closures are
low and can not reflect the practical connection ability of them
even if they have high degree. Moreover, for simplicity, the
computation of nodes' attractiveness only consider the one hop
neighborhood of them. In the near future, there are a number of
interesting extensions could be done. Analyzing the effect of
network structure in more large range may model the spatial factor
more precisely and improve the accuracy of evolution prediction.
Moreover, the change trend of interaction intervals may be an
important role in temporal factor modeling if exact temporal
information is available. Finally, unlike the traditional methods,
this paper evisions evolving networks as dynamic systems and
dynamically investigates node pairs' similarity. We hope the method
and the results in this paper can inspire some new network evolution
modeling methods.

\section*{Acknowledgements}\indent
The work was supported partially by the National Natural Science
Foundation of China (Grant No. 61202255), University-Industry
Cooperation Projects of Guangdong Province (Grant No.
2012A090300001) and the Pre-research Project (Grant No.
51306050102). We thank Tao Zhou and JunMing Shao for their advices.
We also thank Xin Li, Yunpeng Xiao and Yuanping Zhang for
enlightening discussions and careful reading of the manuscript. The
authors also wish to thank the anonymous reviewers for their
thorough review and highly appreciate their useful comments and
suggestions.

\section*{References}

\end{document}